\documentclass[twocolumn,aps,showpacs,prl,superscriptaddress,preprintnumbers,amsmath,amssymb]{revtex4-2}
\usepackage{graphicx}
\usepackage{dcolumn}
\usepackage{appendix}
\usepackage{bm}

\usepackage[usenames,dvipsnames]{xcolor}

\usepackage{subfigure}
\usepackage{bbm}
\usepackage{enumerate}

\usepackage[
  bookmarks=true,
  colorlinks,
  linkcolor=black,
  urlcolor=black,
  citecolor=black,
  plainpages=false,
  pdfpagelabels,
  final,
  breaklinks=true
]{hyperref}

\usepackage{titlesec}

\usepackage{array}
\usepackage{dsfont}

\usepackage{physics}

\begin{document}

\title{High photon number entangled states and coherent state superposition from the extreme-ultraviolet to the far infrared}

\author{Philipp Stammer}
\email{philipp.stammer@icfo.eu}
\affiliation{ICFO -- Institut de Ciencies Fotoniques, The Barcelona Institute of Science and Technology, 08860 Castelldefels (Barcelona), Spain}
\affiliation{Max Born Institute for Nonlinear Optics and Short Pulse Spectroscopy, Max Born Strasse 2a, D-12489 Berlin, Germany}

\author{Javier Rivera-Dean}
\affiliation{ICFO -- Institut de Ciencies Fotoniques, The Barcelona Institute of Science and Technology, 08860 Castelldefels (Barcelona), Spain}

\author{Theocharis Lamprou}
\affiliation{Foundation for Research and Technology-Hellas, Institute of Electronic Structure \& Laser, GR-70013 Heraklion (Crete), Greece}
\affiliation{Department of Physics, University of Crete, P.O. Box 2208, GR-71003 Heraklion (Crete), Greece}

\author{Emilio Pisanty}
\affiliation{Max Born Institute for Nonlinear Optics and Short Pulse Spectroscopy, Max Born Strasse 2a, D-12489 Berlin, Germany}

\author{Marcelo F. Ciappina}
\affiliation{Physics Program, Guangdong Technion--Israel Institute of Technology, Shantou, Guangdong 515063, China}
\affiliation{Technion -- Israel Institute of Technology, Haifa, 32000, Israel}

\author{Paraskevas Tzallas}
\affiliation{Foundation for Research and Technology-Hellas, Institute of Electronic Structure \& Laser, GR-70013 Heraklion (Crete), Greece}
\affiliation{ELI-ALPS, ELI-Hu Non-Profit Ltd., Dugonics tr 13, H-6720 Szeged, Hungary}

\author{Maciej Lewenstein}
\affiliation{ICFO -- Institut de Ciencies Fotoniques, The Barcelona Institute of Science and Technology, 08860 Castelldefels (Barcelona), Spain}
\affiliation{ICREA, Pg. Llu\'{\i}s Companys 23, 08010 Barcelona, Spain}

\date{\today}

\begin{abstract}

We present a theoretical demonstration on the generation of entangled coherent states and of coherent state superpositions, with photon numbers and frequencies orders of magnitude higher than those provided by the current technology. This is achieved by utilizing a quantum mechanical multimode description of the single- and two-color intense laser field driven process of high harmonic generation in atoms. It is found that all field modes involved in the high harmonic generation process are entangled, and upon performing a quantum operation, leads to the generation of high photon number optical cat states spanning from the far infrared to the extreme-ultraviolet spectral region.
This provides direct insights into the quantum mechanical properties of the optical field in intense laser matter interaction.
Finally, these states can be considered as a new resource for fundamental tests of quantum theory, quantum information processing or sensing with non-classical states of light.

\end{abstract}

\maketitle

The superposition of classically distinguishable states is of fundamental interest since the development of quantum theory, and was brought to an extreme by Schrödinger in his famous \textit{Gedankenexperiment} \cite{schrodinger1935gegenwartige}. 
In quantum optics this notion can be retrieved by superpositions of coherent states \cite{gerry1997quantum, ourjoumtsev2007generation}.  
Beside their fundamental interest for testing quantum mechanics \cite{wenger2003maximal, garcia2004proposal}, the generation of these Schrödinger cat states, and of entangled coherent states \cite{sanders1992entangled, wang2001multipartite}, is also of direct technological importance. These states are a powerful tool in quantum information processing \cite{gilchrist2004schrodinger, ourjoumtsev2006generating, vlastakis2013deterministically, jouguet2013experimental}, quantum computation \cite{lloyd1999quantum, ralph2003quantum}, quantum metrology \cite{joo2011quantum}, or can be used to visualize the classical-to-quantum transition \cite{zavatta2004quantum}.
To generate superpositions of coherent states, atom-light interaction in cavities \cite{brune1992manipulation, hacker2019deterministic}, or conditioning approaches at the output of a beam-splitter \cite{dakna1997generating, lund2004conditional, takeoka2007conditional} can be employed. Such conditioning experiments are of general interest in quantum information theory due to their ability for generating entangled optical states \cite{fiuravsek2002conditional}, to describe quantum operations \cite{nielsen2002quantum} or conditional quantum measurements \cite{aharonov1964time, stammer2020state}.  
But, the size of the generated superpositions of coherent states is limited to the range of a few photons, corresponding to moderately small coherent state amplitudes \cite{ourjoumtsev2007generation, takahashi2008generation, mikheev2019efficient}, restricting their applicability in quantum information processing. 
However, due to the relevance of such non-classical states of light in quantum technologies \cite{gilchrist2004schrodinger, ralph2003quantum, walmsley2015quantum}, it is of particular interest to generate a superposition, and entanglement, of coherent states with high photon numbers.
In the present manuscript we show how both can be achieved by means of a conditioning procedure, performed on a so far unrelated photonic platform, namely intense laser-matter interaction. 
Laser sources can easily reach intensities up to $10^{14} \operatorname{W}/\operatorname{cm}^2$, and the field induced material response can be highly non-linear \cite{brabec2000intense}. 
Since these laser fields naturally involve very high photon numbers with corresponding coherent state amplitudes in the range of $\abs{\alpha} = 10^6$, it will be of great advantage to use them for the generation of the sought high photon number non-classical field states \cite{lamprou2020perspective, lewenstein2020quantum}. 
Until recently the intense laser-matter interaction was mainly described by a semi-classical theory, in which the laser field was considered classically such that the properties of the quantum state of the field were not envisioned.  
A commonly used intense laser driven process is the generation of high-order harmonics, in which the coherent properties of the driving laser are transferred to an electronic wavepacket, and later returned to the field modes by the emission of coherent radiation at frequencies of integer multiple of the driving laser field \cite{lewenstein1994theory}. 
However, the recent advances in the quantum optical description of high harmonic generation (HHG) \cite{kominis2014quantum, gonoskov2016quantum, tsatrafyllis2017high, tsatrafyllis2019quantum, lewenstein2020quantum, gorlach2020quantum} allows to conceive new experiments, in which non-classical properties of the field can be observed with the prospective use for modern quantum technologies. 
In particular, it was shown that a conditioning procedure on HHG can lead to non-classical optical Schrödinger cat states in the infrared spectral range \cite{lewenstein2020quantum}. 
To extend the approach to different spectral regions, and to unravel the entanglement between all field modes participating in the HHG process, we have developed a complete quantum mechanical multimode approach. This is used for the description of the interaction of atoms with single- and two-color intense laser fields. We show that all field modes involved in the HHG process are naturally entangled, and upon performing a quantum operation leads to the generation of high photon number coherent state superposition spanning from the extreme-ultraviolet (XUV) to the far-infrared (IR).
We provide the conditions for the generation of XUV cat states, and for the generation of entangled coherent states between two frequency modes in the IR regime with very high photon numbers.

For the description we consider an uncorrelated state prior to the laser-matter interaction $\ket{g} \otimes \ket{\phi}$, in which the atom is prepared in its ground state $\ket{g}$ and the field is described by $\ket{\phi} = \ket{\alpha} \otimes \ket{\{0_q\}}$, where $\ket{ \{ 0_q \}} = \bigotimes_q \ket{0_q}$. The intense driving laser in the fundamental mode is in a coherent state $\ket{\alpha}$, and the harmonic modes $q \in \{2, \, ... \, ,N \} $ are in the vacuum, where the generated harmonics extend to a cutoff $N$. If the interaction is conditioned on the atomic ground state (leading to HHG) \cite{lewenstein1994theory, lewenstein2020quantum}, and neglecting the correlations of the atomic dipole moment \cite{sundaram1990high}, the effective interaction is described by a multimode displacement operator \cite{lewenstein2020quantum, rivera2021quantum}, $D(\chi) = \prod_{q=1}^{N} D(\chi_q)$, where $\chi_q = - i \kappa \sqrt{q} \expval{d}(q \omega)$, with coupling constant $\kappa$, and the Fourier transform of the time-dependent dipole moment expectation value $\expval{d}(q \omega) = \int_{-\infty}^\infty dt \expval{d}(t) e^{i q \omega t}$.
Accordingly, the state of the field after the interaction is shifted $\ket{\phi^\prime} = D(\chi) \ket{\phi} = \ket{\alpha + \chi_1} \otimes_{q=2}^{N} \ket{\chi_q}$. The shift of the coherent state amplitude of the driving laser $\chi_1 = \delta \alpha$ accounts for the depletion of the fundamental mode due to HHG, which are displaced by $\chi_q$. 
However, since the depletion of the fundamental mode and the shift of the harmonic modes are correlated, the actual mode which is excited due to the interaction with the atomic medium is given by a wavepacket mode consisting of all field modes participating in the process. The excitation of this wavepacket mode can be described by the creation operator $B^\dagger$, with the corresponding number states $\ket{ \tilde n}$ satisfying $B^\dagger B \ket{\tilde n} = \tilde n \ket{\tilde n }$. It is this wavepacket mode which is excited during the HHG process \cite{lewenstein2020quantum}. 
In order to take into account the correlation between the shift of the fundamental and harmonic modes, we represent the total state $\ket{\phi^\prime}$ in terms of the  wavepacket mode $\ket{\tilde n}$. Considering only those cases where an excitation of the wavepacket mode is present, but without discriminating between the number of excitation, we project on $\sum_{\tilde n \neq 0} \dyad{\tilde n}$, and obtain 
\begin{align}
\label{eq:projection_wavepacket}
\ket{\psi} = \left[ \mathds{1} - \dyad{\tilde 0} \right] \ket{\alpha + \delta \alpha} \otimes_{q=2}^N \ket{  \chi_q   }.
\end{align}

Recalling that the vacuum state of this wavepacket mode is given by the initial state before the interaction, i.e. $\ket{\tilde 0} \equiv D(\alpha)\ket{0} \otimes \ket{\{0_q\}}$, the total state of the field after the HHG process is given by (up to normalization)
\begin{align}
\label{eq:finalstate}
\ket{\psi} = & \ket{\alpha + \delta \alpha} \otimes_{q=2}^N \ket{\chi_q} \\
& - \bra{\alpha} \ket{\alpha + \delta \alpha} \ket{\alpha} \otimes_{q=2}^N \bra{0_q} \ket{\chi_q} \ket{0_q}. \nonumber 
\end{align}
It shows that in the process of HHG all field modes, including the fundamental and all harmonic modes, are naturally entangled. Thus, whenever harmonics are generated the state of the total optical field is entangled. 
Note that the field modes are entangled in such a way that measuring one mode can leave the entanglement of the other modes intact, which suggest the ability of using HHG for generating high dimensional optical cluster states \cite{raussendorf2001one} which are used for measurement based quantum computation \cite{browne2005resource, walther2005experimental}.

\begin{figure}
    \centering
	\includegraphics[width = 1\columnwidth]{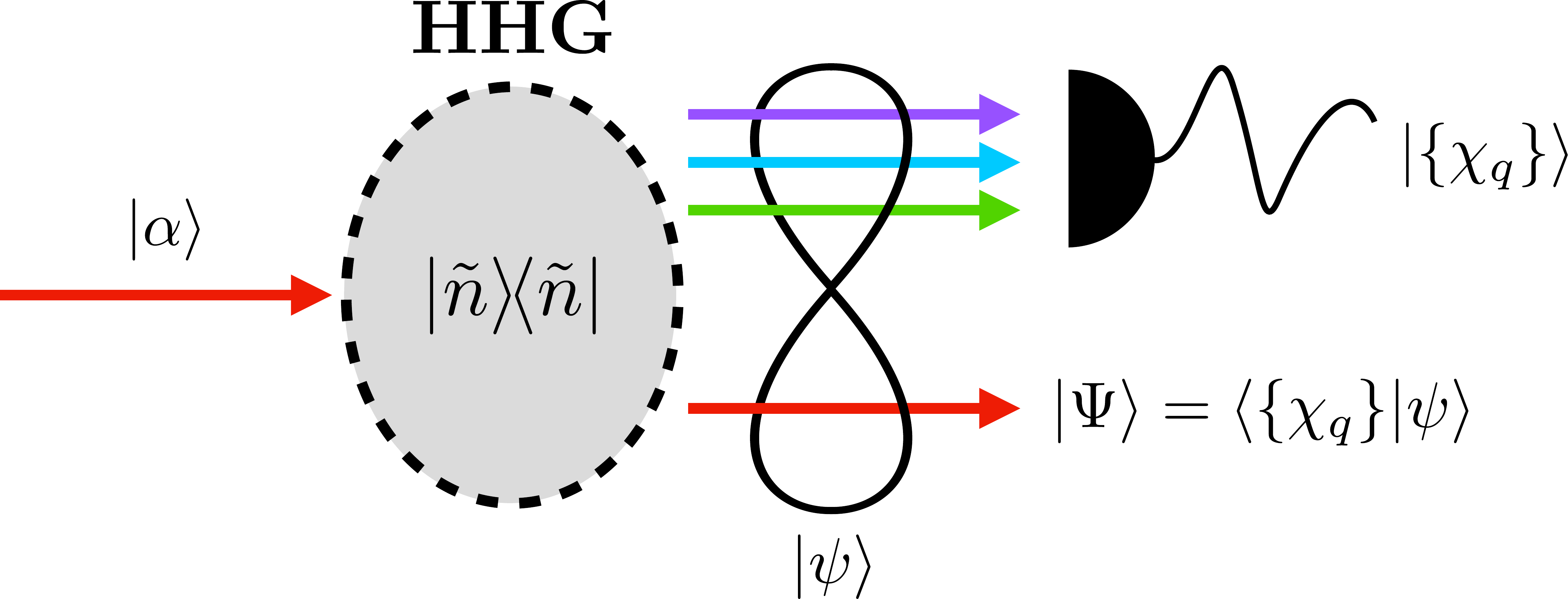}
	\caption{Schematic illustration of the conditioning measurement performed in HHG to generate coherent state superposition. The intense driving laser in the fundamental mode described by the coherent state $\ket{\alpha}$ is interacting with a HHG medium, and the generated optical field is in the wavepacket mode corresponding to $\ket{\tilde n}$. This gives rise to an entangled state between all field modes $\ket{\psi}$. Performing a conditioning measurement on the harmonics modes, by projecting on $\ket{ \{ \chi_q \}}= \bigotimes_q \ket{\chi_q}$, the fundamental mode is found in the coherent state superposition $\ket{\Psi}$.}
      \label{fig:experiment}
\end{figure}

This entangled state, which is heralded by the generation of harmonic radiation, shall now be used to generate non-classical coherent state superposition from the far-IR to XUV spectral region. This is achieved by using the scheme developed in \cite{lewenstein2020quantum} for the generation of optical cat states in the IR regime. 
This scheme is illustrated in Fig. \ref{fig:experiment}, and relies on a post-selection procedure by performing a measurement on the harmonic modes without photon number resolving detectors \cite{lewenstein2020quantum, tsatrafyllis2017high}. 
Part of this measurement constitutes a conditioning on HHG by separating it from other processes (like ionization), i.e. taking only into account the wavepacket excitations via $\ket{\tilde n}$.
Thus, the fundamental mode conditioned on the harmonic signal is given by projecting on the harmonic coherent states $\ket{\Psi} =  \bra{ \{ \chi_q \} } \ket{\psi}$ of amplitude $\chi_q$. The fundamental mode, up to normalization, is then found to be in a superposition of coherent states
\begin{align}
\label{eq:cat}
\ket{\Psi} = \ket{\alpha + \delta \alpha} - \bra{\alpha}\ket{\alpha + \delta \alpha} e^{- \Omega} \ket{\alpha}, 
\end{align}
where $\Omega = \sum_{q > 1} \abs{\chi_q}^2$. 
This state coincides with the state recently reported and measured in \cite{lewenstein2020quantum}, but with the proper prefactor for the second term which takes into account all modes appearing in the experiment. 
To understand the influence of the decoherence factor $\Omega$ we only consider energy conserving events during HHG, i.e. $\abs{\delta \alpha}^2 = \sum_{q=3}^N q \abs{\chi_q}^2$. Assuming that the shift of the harmonics are equal, and using that in single-color HHG only odd harmonics are generated, we find that $\Omega = 2 \abs{\delta \alpha}^2 (N-1)/(N^2+ 2N - 3) $. We thus observe that the influence of the harmonics scales as $\mathcal{O}(1/N)$ with the harmonic cutoff. Due to the extension to large harmonic orders, it makes the scheme intrinsically robust against this kind of decoherence (see Supplementary Material (SM) for more details).

However, due to the complete multimode description of the HHG process developed in this work, we can further generalize this scheme to generate non-classical optical states in extreme wavelength regimes. 
In fact the process of HHG allows to generate entanglement between different frequency modes of the optical field ranging from the far-IR to the XUV regime. Depending on the particular modes measured on \eqref{eq:finalstate}, for instance measuring all modes except $\tilde q \in \{q_i, \, q_j \}$, we obtain the entangled state 
\begin{align}
    \ket{\Psi_{ij}} = \bigotimes_{\tilde q} \ket{\chi_{\tilde q}} - e^{-\Omega_{ij}} \bigotimes_{\tilde q} \bra{0_{\tilde q}} \ket{\chi_{\tilde q}} \ket{0_{\tilde q}}, 
\end{align}
where $\Omega_{ij} = \sum_{q \neq \tilde q} \abs{\chi_{\tilde q}}^2$. Note that for the driving laser mode we have $\ket{0_1} \equiv \ket{\alpha}$ and $\ket{\chi_1} \equiv \ket{\alpha + \delta \alpha}$. This scheme therefore leads to entangled states between IR-IR, IR-XUV and XUV-XUV modes by choosing $q_i$ and $q_j$ appropriately.
For instance, by interchanging the role of the fundamental with the harmonics, i.e. measuring the fundamental mode and projecting \eqref{eq:finalstate} on the coherent state $\ket{\alpha + \delta \alpha}$, we obtain the entangled state of all harmonic modes 
\begin{align}
\ket{\Psi_\Omega} = \bigotimes_q \ket{ \chi_q } - e^{- \abs{\delta \alpha}^2} \bigotimes_q e^{- \frac{1}{2} \abs{\chi_q}^2} \ket{ 0_q}.
\end{align} 

\begin{figure}
    \centering
	\includegraphics[width = 1\columnwidth]{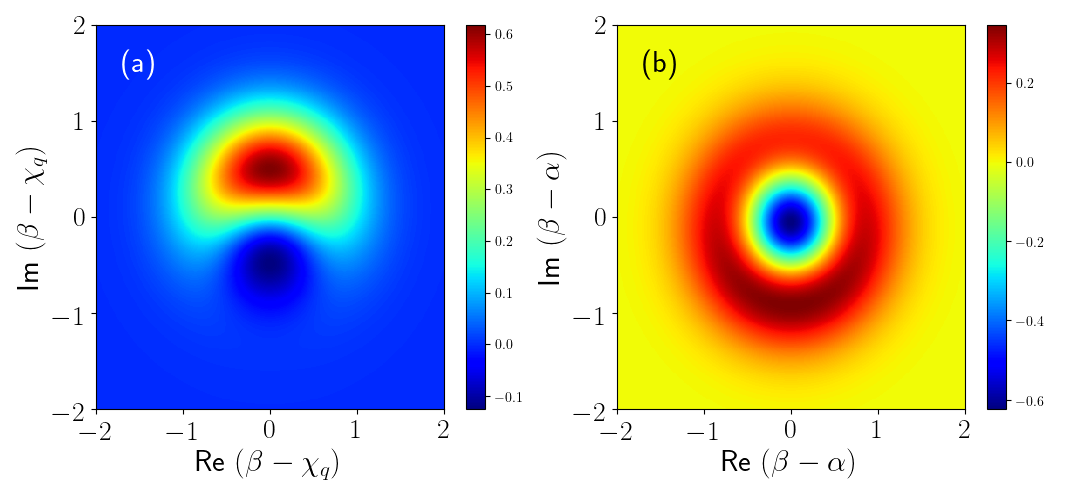}
	\caption{Wigner function of the coherent state superposition (a) of the $q$-th harmonic Eq. \eqref{eq:superposition_harmonic} and (b) of the fundamental mode corresponding to Eq. \eqref{eq:cat}. The calculation has been performed using $\delta \alpha = -0.2$, such that $\chi_q \approx 0.03$ for an harmonic cutoff $N=11$. The opposite shift in imaginary part reflects the correlation between the field modes.}
      \label{fig:wigner_q}
\end{figure}

The fact that the remaining harmonic modes are still entangled, illustrates the peculiar feature of the entangled state in \eqref{eq:finalstate} as an optical cluster state with possible application in quantum information processing.
If we further measure the harmonic modes $q^\prime \neq q$, the state of the $q$-th harmonic is given by 
\begin{align}
\label{eq:superposition_harmonic}
\ket{\Psi_q} = \ket{\chi_q} - e^{- \gamma} \ket{0_q},
\end{align}
where $\gamma = \abs{\delta \alpha}^2+ \Omega^\prime + \frac{1}{2} \abs{\chi_q}^2$ with $\Omega^\prime = \sum_{q^\prime \neq q} \abs{\chi_{q^\prime}}^2$. The state $\ket{\Psi_q}$ represents a superposition of a coherent state with the vacuum in the XUV regime.  
To characterize this state we compute the corresponding Wigner function \cite{royer1989measurement, rivera2021new}
\begin{align}
\label{eq:wigner_q}
W_q (\beta ) = & \frac{2N_q^2}{\pi} \left[ e^{- 2\abs{\beta - \chi_q}^2} + e^{- ( \Omega + \Omega^\prime)} e^{- 2 \abs{\delta \alpha}^2} e^{- 2 \abs{\beta}^2} \right. \nonumber \\
& \left. - e^{- \Omega} e^{- \abs{\delta \alpha}^2} e^{-2 \abs{\beta}^2}  \left( e^{2 \beta \chi_q^*} + e^{2 \beta^* \chi_q}  \right) \right],
\end{align}
with the normalization $N_q$ of \eqref{eq:superposition_harmonic}. 
In Fig. \ref{fig:wigner_q} (a) we show the Wigner function \eqref{eq:wigner_q} for an XUV field of wavelength $\lambda_{XUV}=72.7$ nm for the $11$-th harmonic of a driving laser with frequency $\lambda_{IR}=800$ nm.
For comparison, Fig. \ref{fig:wigner_q} (b) shows the Wigner function of the IR field corresponding to \eqref{eq:cat}.
The non-classical features of the XUV and IR coherent state superposition are clearly visible, as both deviate from the Gaussian distribution of a coherent state and depict negative values. Since the corresponding mean photon number for the Wigner functions of Fig. \ref{fig:wigner_q} (a) is less than one, it makes this scheme an interesting source for generating single XUV photons.

However, to obtain a genuine high photon number coherent state superposition in the XUV regime, a second and independent HHG process can be added to the proposed scheme. In the second HHG process harmonics are generated within the same frequency mode with amplitude ${\chi}_q^\prime$. Formally, by coherently adding the harmonic mode from both schemes, i.e. the low photon number coherent state superposition \eqref{eq:superposition_harmonic} with the high photon number coherent state $\ket{\chi_q^\prime}$, gives rise to the coherent state superposition 
\begin{align}
\label{eq:superposition_harmonic_large}
    \ket{\Psi_q^\prime} = D(\chi_q^\prime) \ket{\Psi_q} = \ket{ \chi_q^\prime + \chi_q} - e^{i \phi^\prime} e^{- \gamma}  \ket{ \chi_q^\prime},
\end{align}
where $\phi^\prime = \operatorname{Im}(\chi_q^\prime \chi_q^*)$. 
Regardless of the small average photon number in \eqref{eq:superposition_harmonic} with $\chi_q \ll 1$, the spatiotemporal overlap with the large amplitude coherent state $\ket{\chi_q^\prime}$ leads to the effective displacement operation in \eqref{eq:superposition_harmonic_large} \cite{paris1996displacement}. 
Thus, the coherent state superposition in the IR \eqref{eq:cat} and XUV \eqref{eq:superposition_harmonic_large} describe large amplitude optical cat states. 
Taking into account the typical photon numbers of the IR driving field and the conversion efficiency of the HHG process \cite{chatziathanasiou2017generation}, the IR and XUV cat state can be produced with photon numbers in the range of $\expval{n_{IR}} \sim10^{13}$ and $\expval{n_{XUV}} \sim 10^{7}$ photons per pulse, respectively. 
Here we want to emphasize that HHG is performed in vacuum such that the optical field is subjected to negligible environmental decoherence effects during propagation. 
In addition to the technological importance of generating coherent state superpositions in the XUV and IR spectral range, they depict a notable feature which has a direct consequence to fundamental test of quantum theory. The opposite shift in the imaginary part of their Wigner function is a result of the correlation in the shift of the coherent state amplitudes of the field modes in the process of HHG. Such quantum correlations between field modes can be used, via homodyne quadrature measurements, towards violating Bell type inequalities \cite{wenger2003maximal, garcia2004proposal}.

However, to generate genuine high photon number entangled coherent states in the order of $\abs{\alpha} = 10^6$ we generalize the HHG process by using a two-color driving field. Such high harmonic generation experiments are often performed for a $\omega-2\omega$ laser frequency configuration with frequencies in the visible to far-infrared spectral region, with parallel or orthogonal polarizations between the two driving lasers \cite{kim2005highly, mauritsson2006attosecond, fleischer2014spin}. In this case, the initial state of the two mode driving field is given by $\ket{\alpha_1} \otimes \ket{\alpha_2}$, such that the total field after the interaction with the HHG medium is given by $\ket{\alpha_1 + \delta \alpha_1} \otimes \ket{\alpha_2 + \delta \alpha_2} \otimes \ket{ \{ \bar{\chi}_q \}}$, where $\delta \alpha_1$ and $\delta \alpha_2$ are the depletion of the two driving field modes, respectively. Following the procedure introduced above, and after taking into account the correlations via the corresponding wavepacket mode, the obtained state reads 
\begin{align}
\ket{\Psi} = &  \ket{\alpha_1 + \delta \alpha_1} \otimes \ket{\alpha_2 + \delta \alpha_2} \bigotimes_{q>2} \ket{\bar{\chi}_q}   \\
&  - e^{- i \left( \varphi_1 + \varphi_2\right)} e^{- \frac{1}{2} \Delta} \ket{\alpha_1} \otimes \ket{\alpha_2} \bigotimes_{q>2} e^{- \frac{1}{2} \abs{\bar{\chi}_q}^2} \ket{0_q} , \nonumber
\end{align}
where $\varphi_i = \operatorname{Im}(\alpha_i \delta \alpha_i^*)$ and $\Delta = \abs{\delta \alpha_1}^2 + \abs{\delta \alpha_2}^2$. By conditioning on the harmonic signal, i.e. projecting on $\ket{\{ \bar{\chi}_q\}}$, we obtain
\begin{align}
\label{eq:ECS}
\ket{ECS} = & \ket{\alpha_1 + \delta \alpha_1} \otimes \ket{\alpha_2 + \delta \alpha_2} \\
&  -  e^{- i \left( \varphi_1 + \varphi_2\right)} e^{- \frac{1}{2} \Delta} e^{- \bar \Omega} \ket{\alpha_1} \otimes \ket{\alpha_2}, \nonumber
\end{align}
with $\bar \Omega =\sum_{q>2} \abs{\bar{\chi}_q}^2$.
This scheme can be utilized in the spectral range from the visible to the far-infrared regime, which is within the telecom optical fiber wavelength regime useful for long distance entanglement distribution for quantum information processing due to minimized attenuation.
Due to the high degree of coherent control in the two-color HHG processes, the relative field amplitudes and phase of the amplitude entangled state between the two physical frequency modes can be tailored in a controllable way. For instance by independently varying the driving field amplitudes $\alpha_i$ or the relative depletion $\delta\alpha_i $ via the respective field polarization, e.g. linear or circular orthogonal polarized fields. 

\begin{figure}
    \centering
	\includegraphics[width = 1\columnwidth]{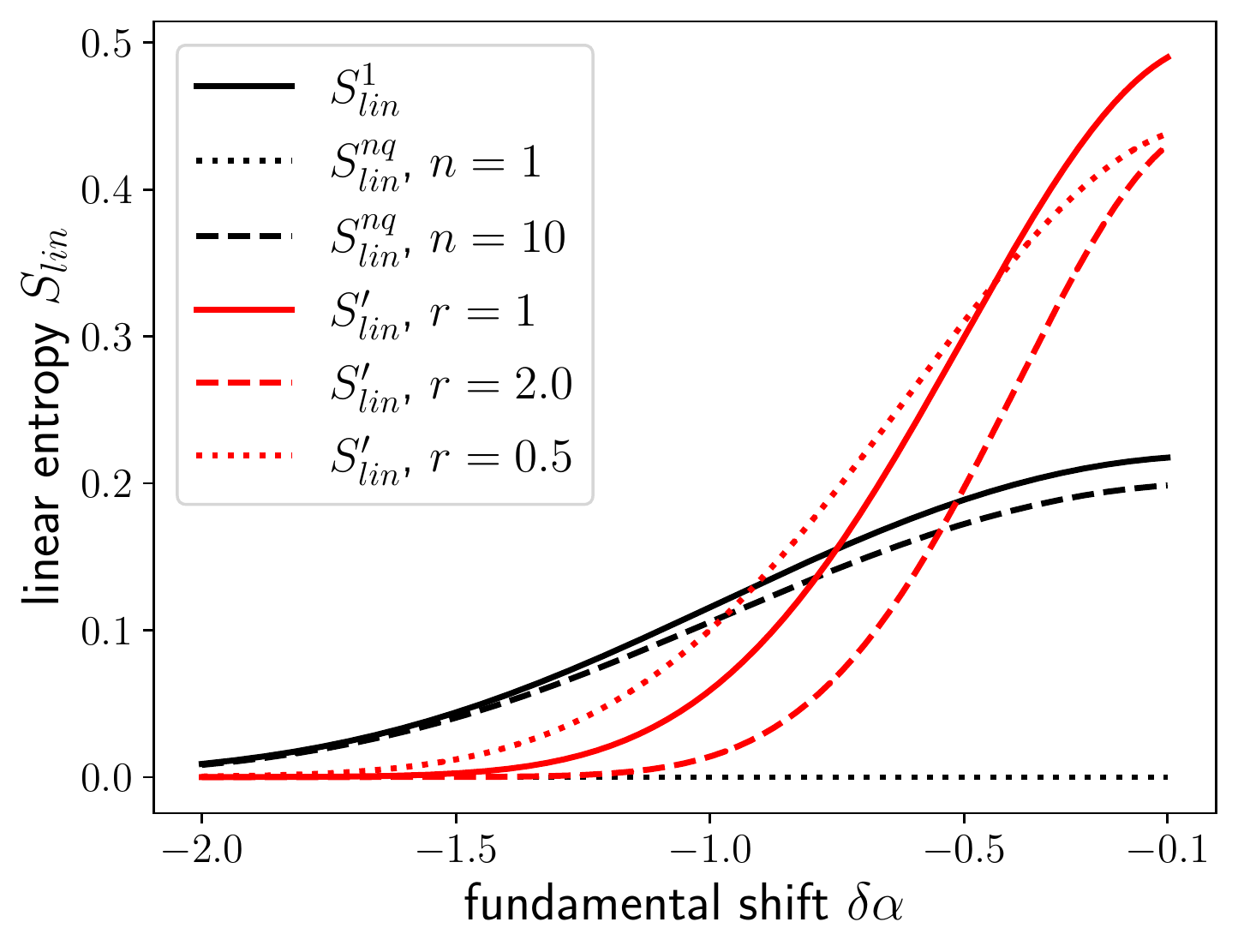}
	\caption{Linear entropy measures $S_{lin}^1$ (black solid) and $S_{lin}^{nq}$ for two different partitions of the entangled state \eqref{eq:finalstate} with $n=1$(black dashed) and $n=10$ (black dotted) for increasing depletion of the fundamental mode $\abs{\delta \alpha}$. The linear entropy measure for the two-color high harmonic generation experiment $S_{lin}^\prime$ (red) with different ratios of the depletion of the two driving fields $r = \abs{\delta \alpha_2}^2 / \abs{\delta \alpha_1}^2 $. In all cases we have used the harmonic cutoff at $N = 11$. }
      \label{fig:entanglement}
\end{figure}

In order to compare the single- and two-color HHG setup we shall quantify the degree of entanglement between the field modes in each scheme. We will make use of the degree of purity of the reduced density matrix of a subsystem. Since the reduced density matrix of an entangled state is not pure $\rho^2 \neq \rho$, we use the linear entropy $S_{lin} = 1 - \Tr (\rho^2)$ as a quantitative measure of the involved entanglement between coherent states \cite{agarwal2005quantitative, berrada2013beam}. Since $\Tr (\rho^2)  \le 1$, where the equality only holds for pure states, a non-vanishing linear entropy serves as a witness of entanglement in the total system.
For the single-color HHG experiment \eqref{eq:finalstate} we particularly focus on two cases. First, on the entanglement between the fundamental driving field with all harmonic modes, and second, on the entanglement of $n$ harmonic modes with all remaining modes (including the fundamental). We thus compute the reduced density matrices of the fundamental mode, and for the $q \in \{2, \, ..., \, n+1 \} $ harmonics via $\rho_{q=1} = \Tr_{q>1} (\dyad{\psi})$, and $\rho_{nq} = \Tr_{q^\prime \neq q} (\dyad{\psi})$, respectively. The corresponding linear entropy measures, for the fundamental mode $S_{lin}^1$ (black, solid) and for $n$ harmonics $S_{lin}^{nq}$ (black dashed and dotted), are shown in Fig. \ref{fig:entanglement} as a function of $\abs{\delta \alpha}$ (SM). 
The entanglement witness depends on the depletion of the fundamental mode $\delta \alpha$, which is correlated to the harmonic amplitude $\chi_q$.  
We observe that the entanglement between the fundamental mode with the harmonics (solid) is larger than the entanglement between $n$ harmonic modes with all other field modes (dashed, dotted), and that the entanglement measure monotonically decreases for an increasing depletion of the fundamental mode. 
For large $\delta \alpha$, all entanglement decays since the amplitude of the second term in \eqref{eq:finalstate} vanishes due to the decreasing overlap between the two coherent states. 
For the different partitions of the $n$ harmonic modes we observe that for larger $n$ the entanglement with the remaining modes is increased and almost negligible for $n=1$. 
For the two-color HHG process in which an entangled pair of coherent states of large amplitude can be generated \eqref{eq:ECS}, we quantify the involved entanglement by tracing over the second $2 \omega$ driving field mode in \eqref{eq:ECS}. The corresponding linear entropy $S_{lin}^\prime$ is shown in Fig. \ref{fig:entanglement} (red) for different ratios of the depletion of the two modes $r = \abs{\delta \alpha_2}^2 / \abs{\delta \alpha_1}^2$ (SM). We observe that for small depletion the involved entanglement between the two driving field modes in the two-color HHG experiment is larger than in any single-color experiment, and is the largest for equal depletion (red, solid). For a larger depletion $\abs{\delta \alpha}$ the entanglement decays slower when the $2 \omega$ field has smaller depletion than the $\omega$ mode, e.g. $r = 0.5$ (red, dotted). In the opposite case, with $r = 2.0$, the entanglement is smallest (red, dashed). 

In conclusion, we developed a quantum mechanical multimode approach for the description of HHG driven by single- and two-color intense laser fields. We showed that all field modes involved in the HHG process are naturally entangled once harmonics are generated. Performing quantum operations on particular field modes leads to the generation of high photon number coherent state superpositions spanning from the XUV to the far IR spectral region. We provided the conditions for the generation of XUV-IR correlated coherent state superposition, and the generation of entangled states in the visible-IR spectral region with controllable quantum features.
The entangled states generated by using HHG are deterministically generated whenever harmonic radiation is emitted, and the coherent state superposition is heralded when the conditioning measured is performed. 
Considering that similar HHG mechanism underlie the majority of the intense-laser matter interactions \cite{osika2017wannier, tsatrafyllis2019quantum, lamprou2021quantum}, we anticipate that the findings will set the stage for conceiving novel experiments for the generation of a whole family of high photon number non-classical entangled field states, challenging the quantum state characterization schemes \cite{fuchs2022photon}, advancing fundamental studies of quantum theory and provide a new platform for optical quantum technologies \cite{walmsley2015quantum}. 
Finally, we note that the dynamics of the HHG process is intrinsically in the attosecond time regime, which further stress the potential impact of the present work on quantum information technologies towards a previously inaccessible time scale, and can further be used for optical signaling and spectroscopy with non-classical light states \cite{dorfman2016nonlinear}.

\begin{acknowledgments}

ICFO group acknowledges support from ERC AdG NOQIA, from Agencia Estatal de Investigaciín (the R$\&$D project CEX2019-000910-S, funded by MCIN/ AEI/10.13039/501100011033, Plan National FIDEUA PID2019-106901GB-I00, FPI, QUANTERA MAQS PCI2019-111828-2, Proyectos de I+D+I “Retos Colaboración” RTC2019-007196-7) from Fundació Cellex, Fundació Mir-Puig, and from Generalitat de Catalunya through the CERCA program, AGAUR Grant No. 2017 SGR 134, QuantumCAT \ U16-011424, co-funded by ERDF Operational Program of Catalonia 2014-2020), EU Horizon 2020 FET-OPEN OPTOLogic (Grant No 899794), and the National Science Centre, Poland (Symfonia Grant No. 2016/20/W/ST4/00314), Marie Sk\l odowska-Curie grant STREDCH No 101029393, “La Caixa” Junior Leaders fellowships (ID100010434), and EU Horizon 2020 under Marie Sk\l odowska-Curie grant agreement No 847648 (LCF/BQ/PI19/11690013, LCF/BQ/PI20/11760031, LCF/BQ/PR20/11770012).).
FORTH group acknowledges LASERLABEUROPE (H2020-EU.1.4.1.2 Grant ID 654148),
FORTH Synergy Grant AgiIDA (Grand No. 00133), the EU’s H2020 framework programme
for research and innovation under the NFFA-Europe-Pilot project (Grant No. 101007417).
J.R-D. acknowledges support from the Secretaria d'Universitats i Recerca del Departament d'Empresa i Coneixement de la Generalitat de Catalunya, as well as the European Social Fund (L'FSE inverteix en el teu futur)--FEDER.

\end{acknowledgments}

\bibliographystyle{unsrt}
\bibliography{references}{}

\end{document}


\title{Supplementary Material: High photon number entangled states and coherent state superposition from the extreme-ultraviolet to the far infrared}

\date{\today}

\maketitle





\section{\label{sec:appendix_econservation}Energy conservation during high harmonic generation}

By considering only the energy conserving events in the HHG process, i.e. in which the energy of the driving laser field is transferred to the harmonic field modes only, and no additional processes take place, we have to fulfill 
\begin{align}
\abs{\delta \alpha}^2 = \sum_{q=3}^N q \abs{\chi_q}^2,
\end{align}
where we have neglected the contributions from the vacuum fluctuations. To evaluate the contributions of the harmonic modes on the right hand side, we assume that the shift of the harmonic modes are equal $\abs{\chi_q}^2 = \abs{\chi}^2$, which is fulfilled in HHG due to the plateau structure of the high harmonic intensities. Further using that only odd harmonics are generated for single color driving fields, we can write 
\begin{align}
\abs{\chi}^2 \sum_{q=3}^N q = \abs{\chi}^2 \frac{N^2 + 2N - 3}{4}. 
\end{align}

Using that $\Omega  = \sum_{q=3}^N \abs{\chi_q}^2 = \abs{\chi}^2(N-1)/2$, for the odd harmonics, we have
\begin{align}
\label{eq:relation_energyconservation}
\Omega = \frac{2(N-1)}{N^2+2N-3} \abs{\delta \alpha}^2.
\end{align}

And we observe that the environmental induced decay parameter $\Omega$ becomes smaller for an increased high harmonic cutoff and scales as $\mathcal{O}(1/N)$. Thus high harmonic generation itself stabilizes against this kind of decoherence, since the cutoff usually extends to values $N\ge 15$. 

For the case of a two-color $\omega-2 \omega$ driven HHG process the energy conservation includes the depletion of both driving modes such that 
\begin{align}
\abs{\delta \alpha_1}^2 + 2 \abs{\delta \alpha_2} = \sum_{q=3}^N q \abs{\bar{\chi}_q}^2.
\end{align}

With the same assumption about the plateau structure of the harmonics, i.e. $\abs{\bar{\chi}_q}^2 = \abs{\bar{\chi}}$, the summation on the right hand side is over all harmonics (even and odd) and thus we have 
\begin{align}
\abs{\delta \alpha}^2 ( 1 + 2 r ) = \abs{\bar{\chi}}^2 \frac{N^2+N}{2} - 3 ,
\end{align}
where we have introduced the ratio between the driving laser field amplitudes $r = \abs{\delta \alpha_2}^2/ \abs{\delta \alpha_1}^2 $. Further evaluating $\bar \Omega = \sum_{q=3}^N \abs{\bar{\chi}_q}^2 =\abs{\bar{\chi}}^2 (N-2) $, including even and odd harmonics for a two-color driving laser, we have 
\begin{align}
\label{eq:decoherence_2color}
\bar \Omega = \frac{2N-4}{N^2+N-6} (1+2r) \abs{\delta \alpha_1}^2.
\end{align}

Note that in general the depletion of the fundamental mode $\abs{\delta \alpha}$ itself has a complicated dependence on the harmonic cutoff $N$. 
The dependence is determined by the particular experimental configuration including the driving field intensity, its frequency and polarization. It depends as well on the details of the HHG medium like the gas density or the alignment with respect to the laser polarization axis in case of molecular media. Also other mechanisms may intervene such as the absorption of the generated harmonic radiation within the medium itself.

\section{\label{sec:wigner_dependence_decoherence}Influence of the decoherence factor on the fundamental mode}

\begin{figure}
    \centering
	\includegraphics[scale=0.37]{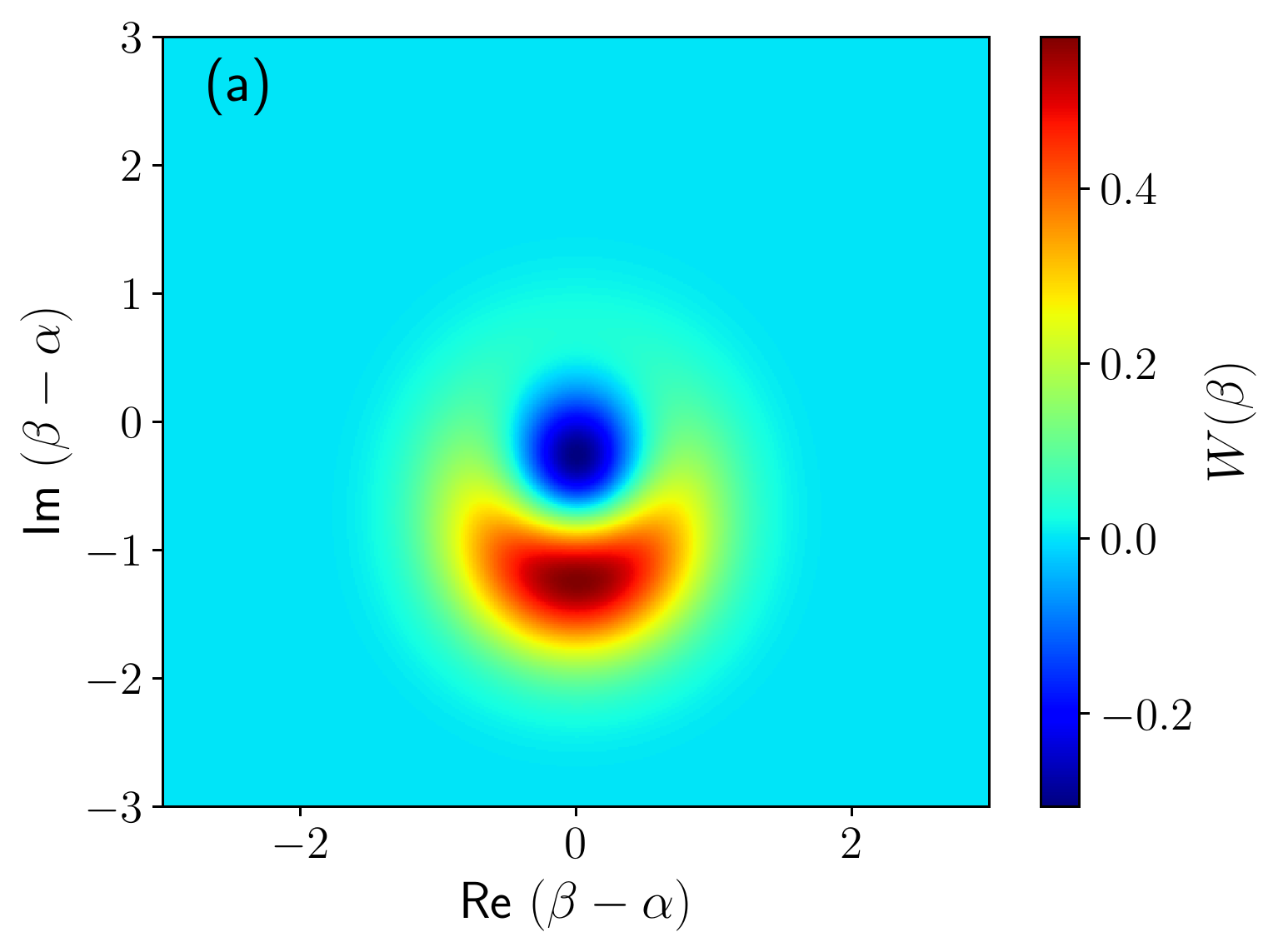}
	\includegraphics[scale=0.37]{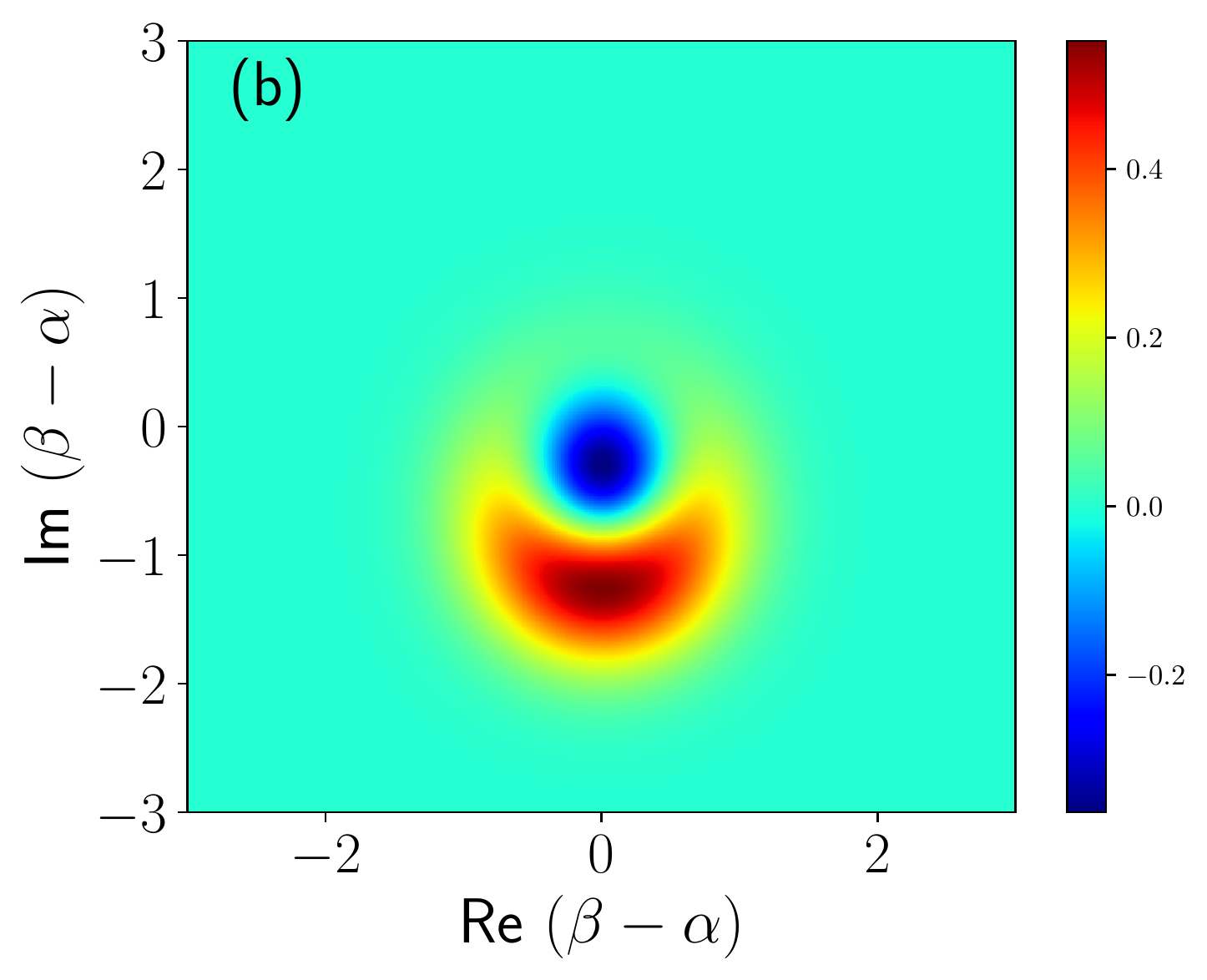}
	\includegraphics[scale=0.37]{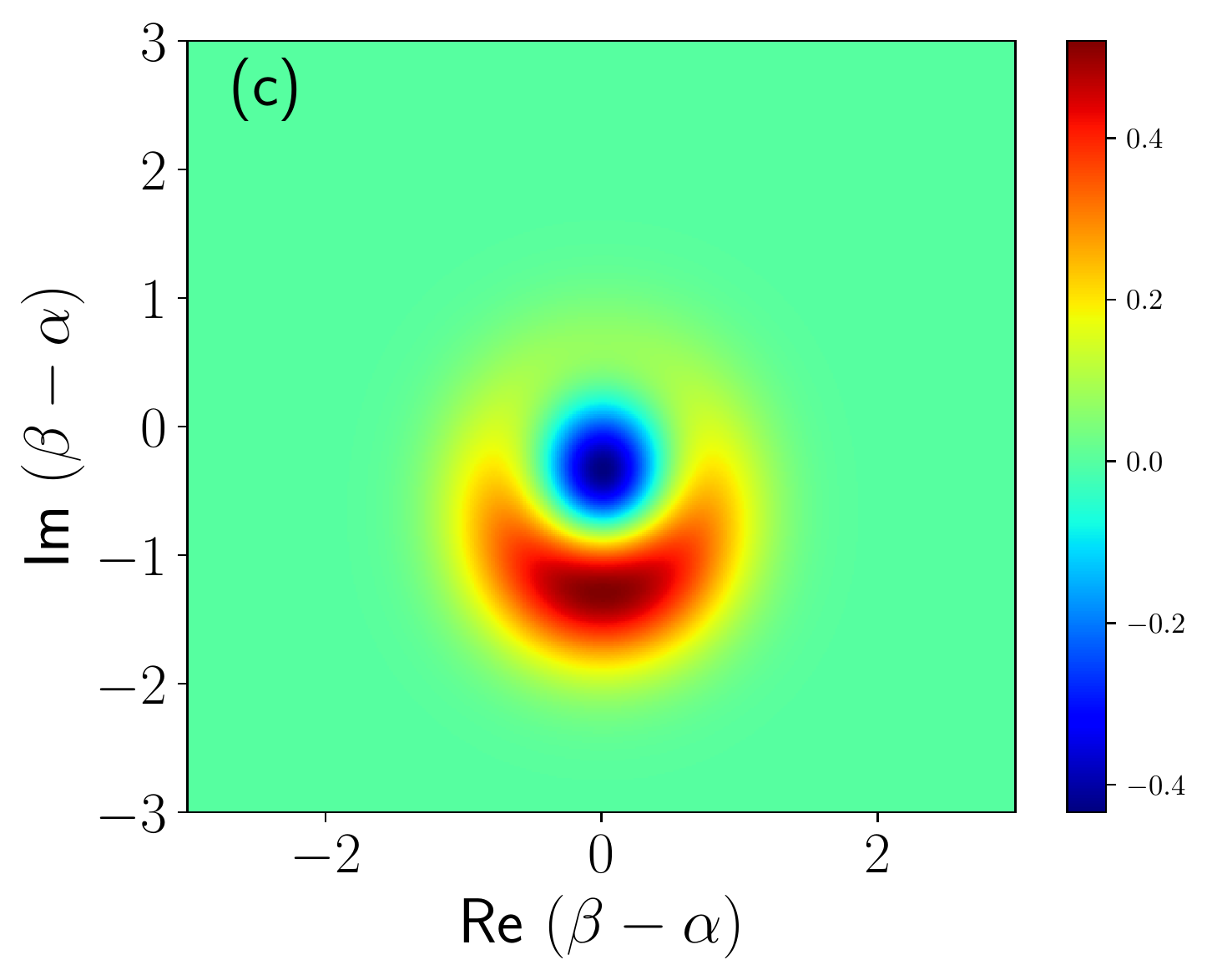}
	\includegraphics[scale=0.37]{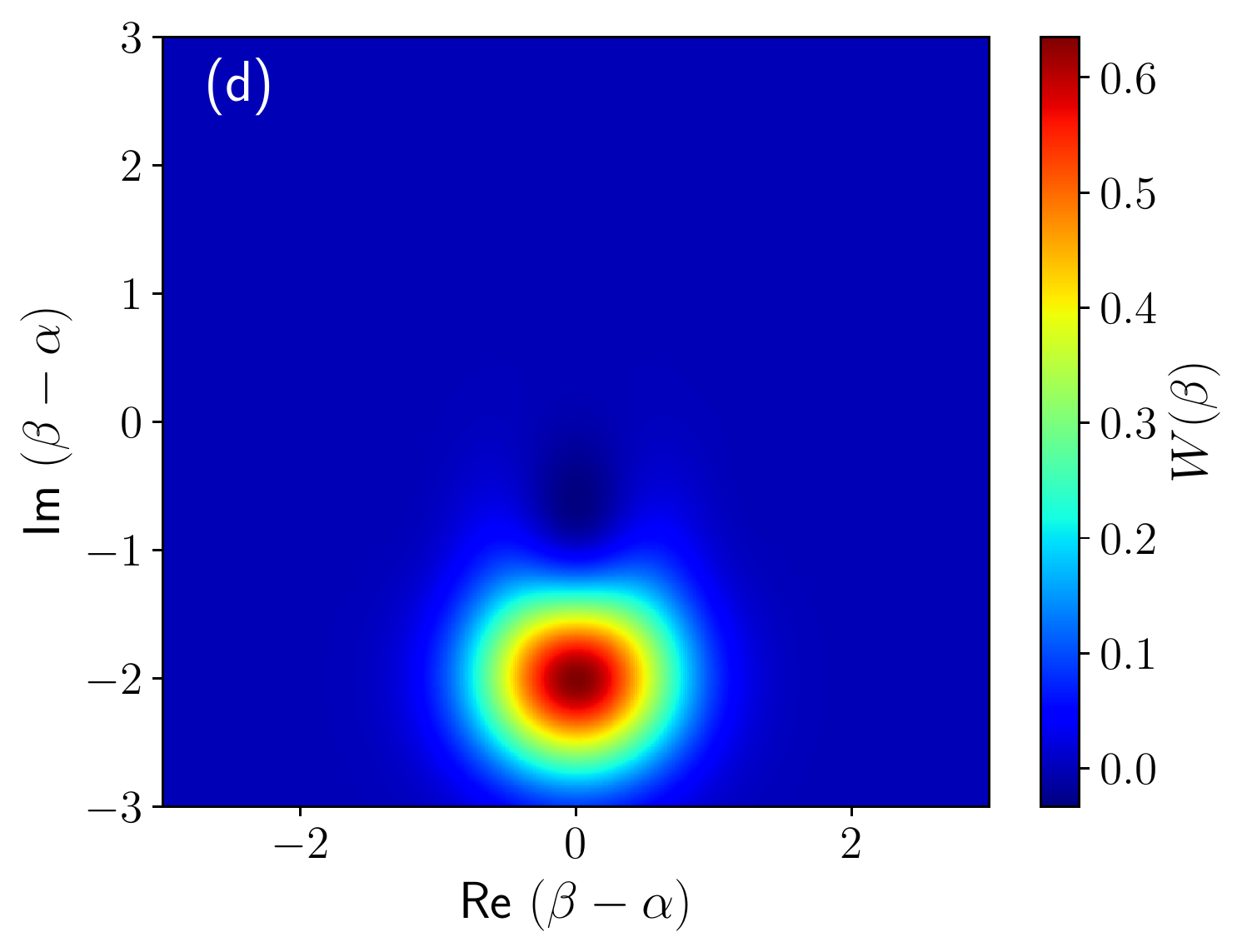}
	\includegraphics[scale=0.37]{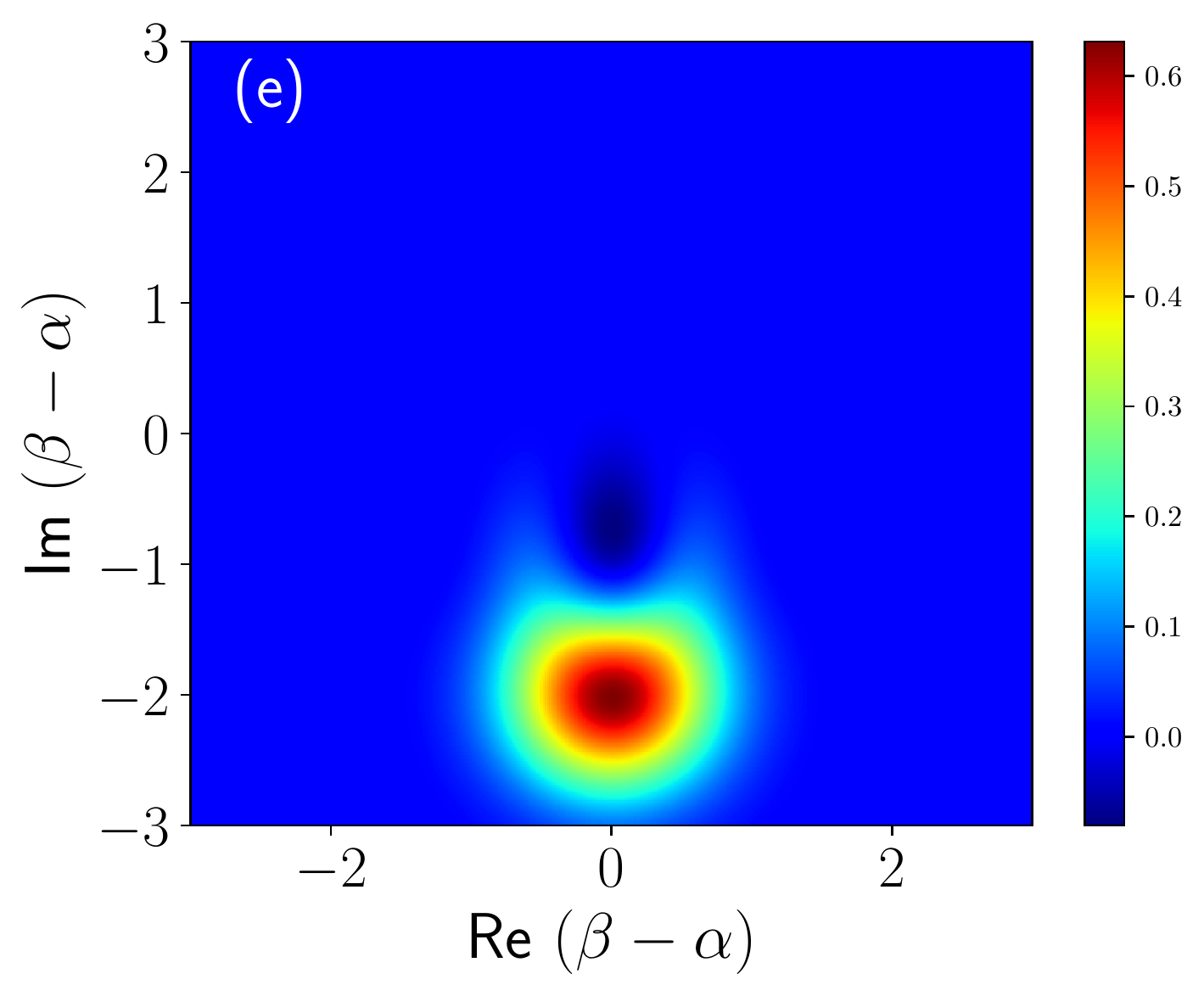}
	\includegraphics[scale=0.37]{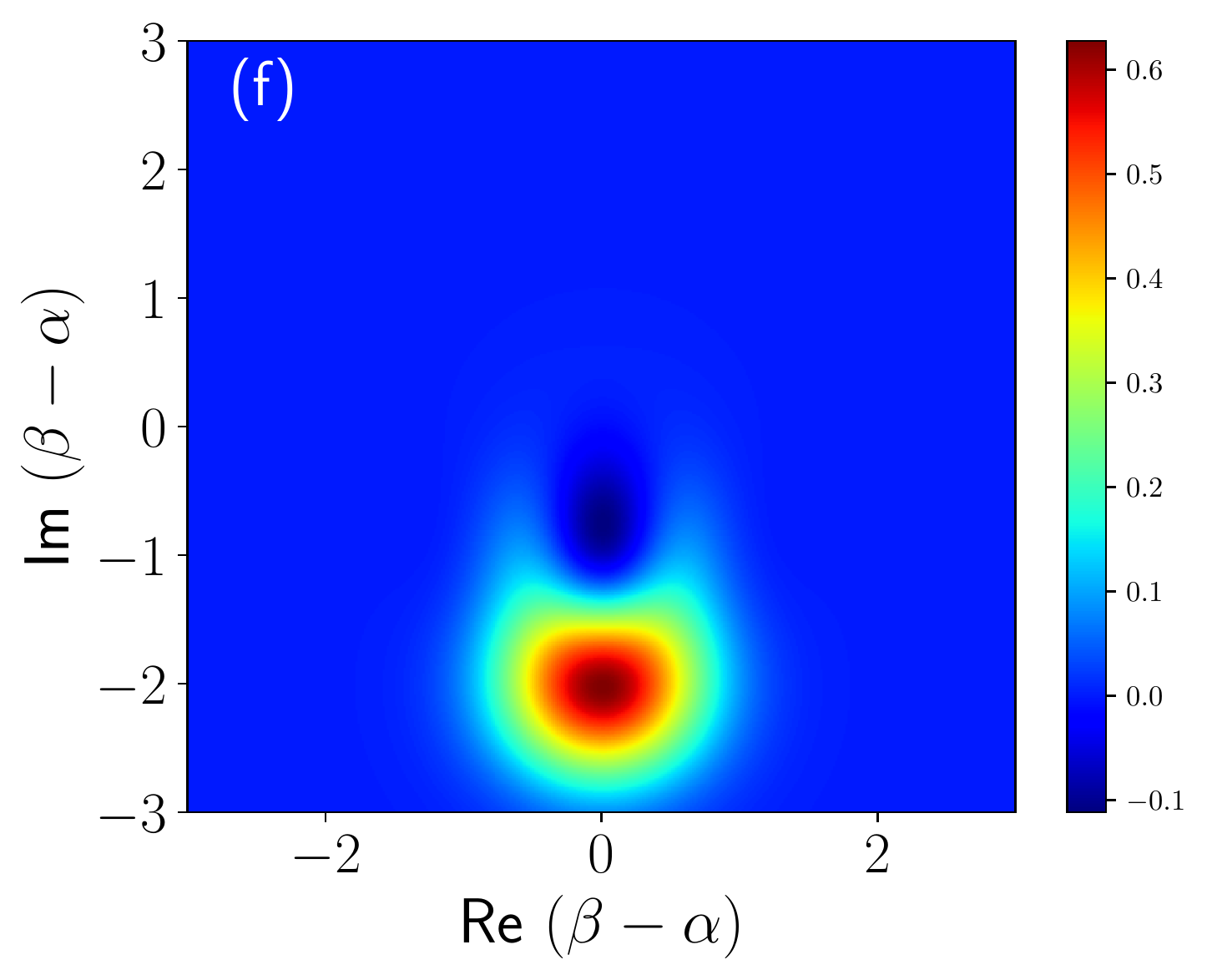}	
	\caption{Wigner function of the coherent state superposition for $\delta \alpha = -1.0$ (top row) and different decoherence factors, (a) $N=5$, (b) $N=11$ and (c) $N=101 $. Wigner function for $\delta \alpha = -2.0$ (bottom row) and (d) $N=5$, (e) $N=21$ and (f) $N=101$.}
      \label{fig:wigner_decoherence}
\end{figure}

To characterize the coherent state superposition of the fundamental mode 
\begin{align}
\label{eq:cat}
\ket{\Psi} = \ket{\alpha + \delta \alpha} - \bra{\alpha}\ket{\alpha + \delta \alpha} e^{- \Omega} \ket{\alpha}, 
\end{align}
we compute the corresponding Wigner function
\begin{align}
\label{eq:wigner}
W(\beta) =&  \frac{2\, \mathcal{N}^2}{\pi} \left[ e^{- 2 \abs{\beta - \alpha - \delta \alpha}^2 } + e^{-\Omega} e^{- \abs{\delta \alpha}^2} e^{- 2 \abs{\beta - \alpha}^2}  \times \left( e^{- \Omega} -  e^{2 (\beta - \alpha) \delta \alpha^*} - e^{2(\beta - \alpha)^* \delta \alpha}   \right) \right], 
\end{align}
with the normalization $\mathcal{N}$ of \eqref{eq:cat}. 
To analyze the influence of the decoherence factor $\Omega$ on the signatures of the Wigner function we consider different high harmonic generation processes by varying the harmonic cutoff $N$. Since the decoherence factor depends on the high harmonic cutoff $N$ according to \eqref{eq:relation_energyconservation}, we show the Wigner function \eqref{eq:wigner} for the different cutoffs $N \in \{ 5, 11, 101 \}$ for $\delta \alpha = -1.0$ in Fig. \ref{fig:wigner_decoherence} (a) - (c). 
We can see that for an increasing high harmonic cutoff $N$, and consequently reduced $\Omega$, the non-classical feature of the Wigner function, by means of the negativity, is enhanced. Since the high harmonic generation process usually extends to high orders, i.e. large $N$, implies that the process stabilizes itself against the influence of the decoherence of the coherent state superposition in the fundamental mode induced by the residual field modes since $\Omega \to 0$ for $N \to \infty$. 
This effect is even more pronounced when the shift of the fundamental mode is increased, which also increases $\Omega$ due to \eqref{eq:relation_energyconservation}. In Fig. \ref{fig:wigner_decoherence} (d) - (f) we show the Wigner function \eqref{eq:wigner} for $\delta \alpha = -2.0$, where the overlap of the second term in the coherent state superposition \eqref{eq:cat} is already strongly reduced and thus almost all negativity vanishes for a small harmonic cut-off in (d) and only rudimentary appear in (e) corresponding to $N=5$ and $N=21$, respectively. However, we can see that for large high harmonic cut-off $N =101$ in (f) the negativity of the Wigner function can be clearly seen and thus s state tomography can show non-classical signatures. This signature is robust for large harmonic cut-off frequencies, and further show the intrinsic stabilization of the generated coherent state superposition in HHG.

\section{\label{sec:appendix_HHG_conditioning}Coherent state superposition in the extreme ultraviolet regime}

We have shown that the process of high harmonic generation with a conditioning procedure can be used to generate non-classical coherent state superpositions in the extreme ultraviolet regime. This was achieved by projecting the entangled state from high harmonic generation 
\begin{align}
\ket{\psi} =  \ket{\alpha + \delta \alpha} \otimes_q \ket{\chi_q} - \bra{\alpha} \ket{\alpha + \delta \alpha} \ket{\alpha} \otimes_q \bra{0_q} \ket{\chi_q} \ket{0_q},  
\end{align}
on the shifted state of the fundamental mode $\ket{\alpha + \delta \alpha}$ and the state $\ket{ \{ \chi_{q^\prime} \}}$ of all harmonics $q^\prime \neq q$, such that the state of the $q$-th harmonic is given by 
\begin{align}
\label{eq:cat_harmonic}
\ket{\Psi_q} = N_q \left[ \ket{\chi_q} - e^{- \gamma} \ket{0_q} \right].
\end{align}
where $N_q = [1+ e^{- \Omega} (e^{- 2 \abs{\delta \alpha}^2} e^{- \Omega^\prime} - 2 e^{- \abs{\delta \alpha}^2} )]^{-1/2}$, $\gamma = \abs{\delta \alpha}^2 + \Omega^\prime + \frac{1}{2} \abs{\chi_q}^2$ and $\Omega^\prime = \abs{\chi}^2 \frac{N-3}{2} = \frac{N-3}{N-1} \Omega$.

The corresponding Wigner function of the coherent state superposition of the $q$-th harmonic \eqref{eq:cat_harmonic} is given by 
\begin{align}
\label{eq:wigner_q}
W_q (\beta ) = & \frac{2N_q^2}{\pi} \left[ e^{- 2\abs{\beta - \chi_q}^2} + e^{- ( \Omega + \Omega^\prime)} e^{- 2 \abs{\delta \alpha}^2} e^{- 2 \abs{\beta}^2}  - e^{- \Omega} e^{- \abs{\delta \alpha}^2} e^{-2 \abs{\beta}^2}  \left( e^{2 \beta \chi_q^*} + e^{2 \beta^* \chi_q}  \right) \right].
\end{align}

\begin{figure}
    \centering
	\includegraphics[scale=0.5]{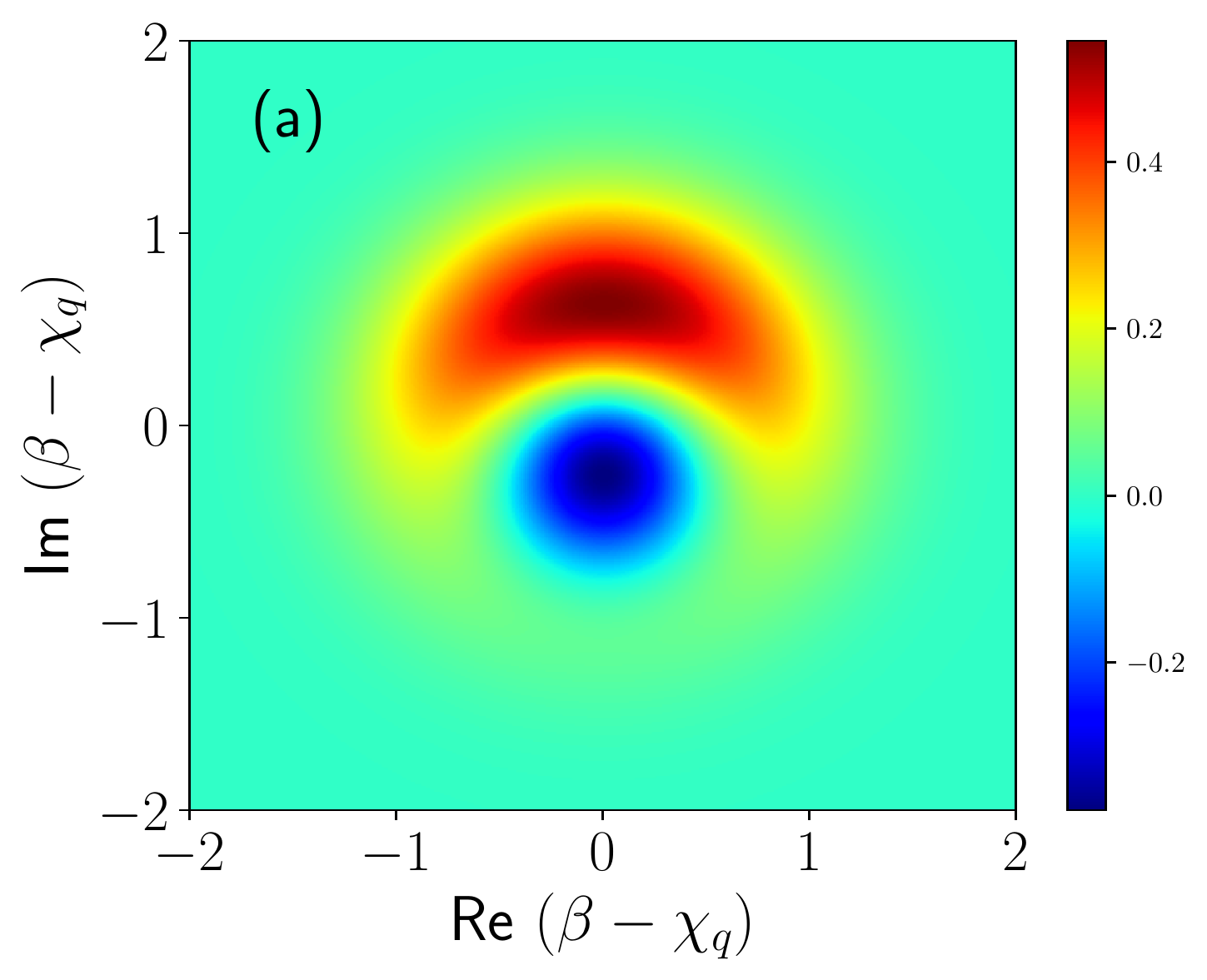}
	\includegraphics[scale=0.5]{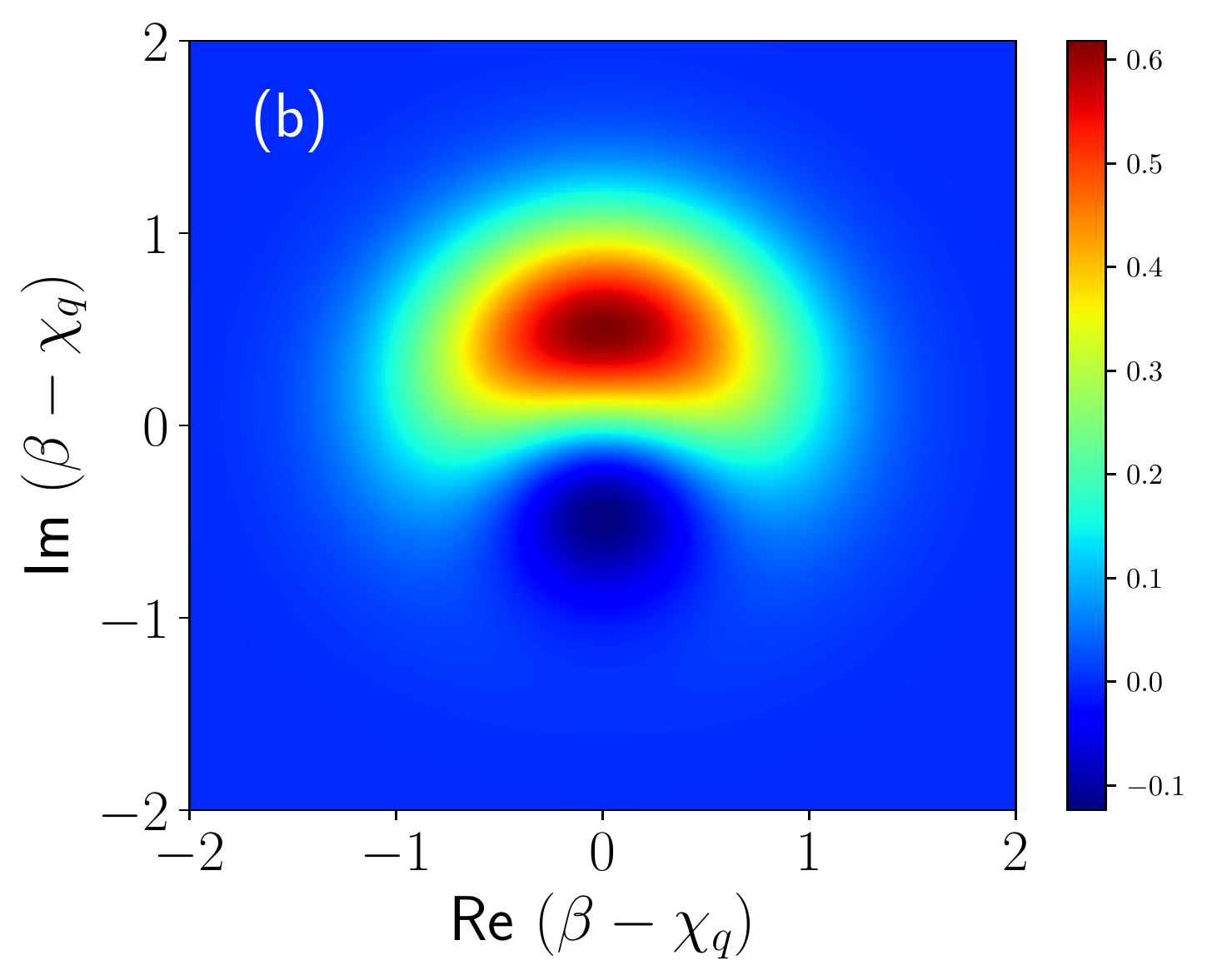}
	\includegraphics[scale=0.5]{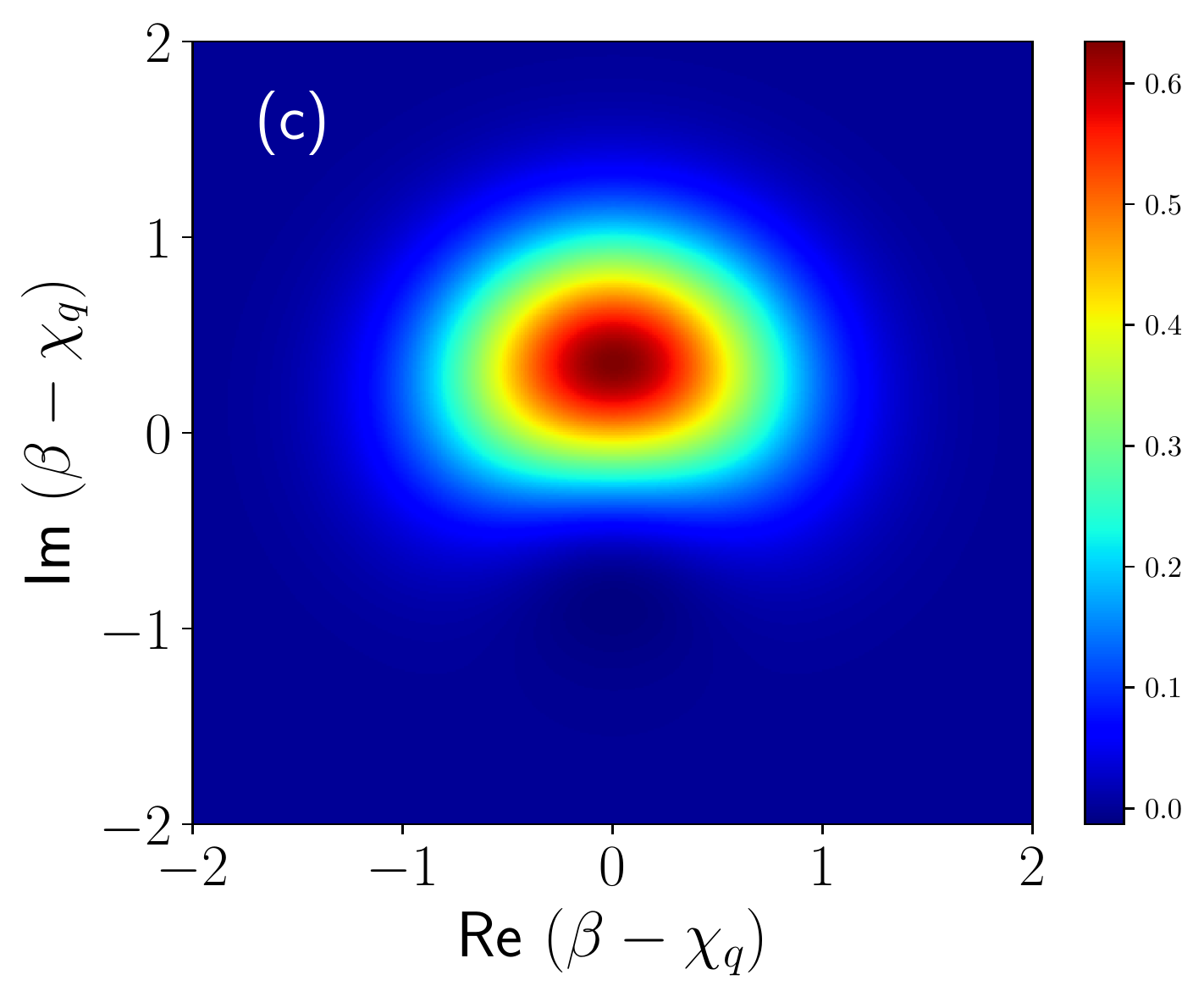}	
	\includegraphics[scale=0.5]{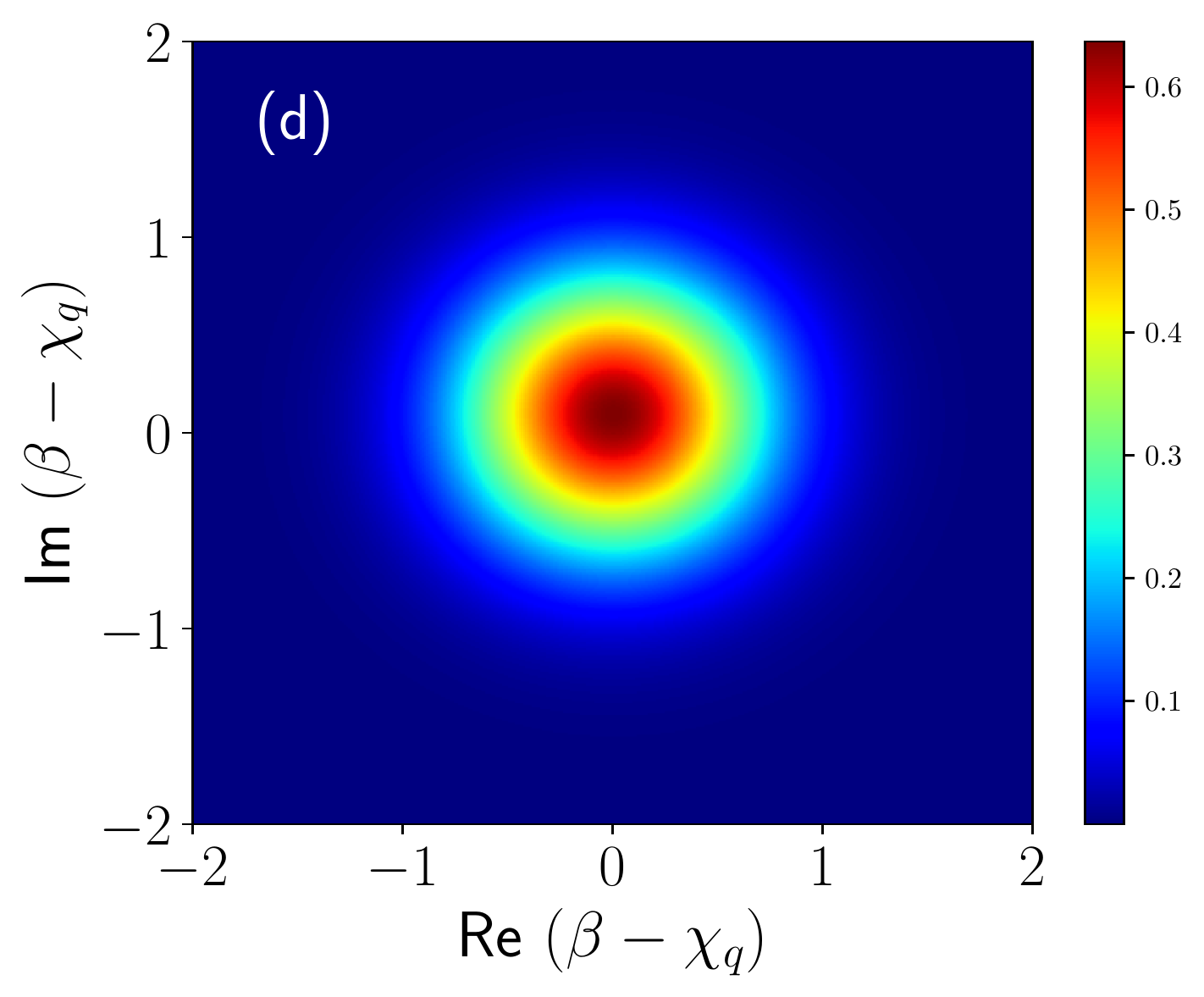}
	\caption{Wigner function of the coherent state superpositions of the $q$-th harmonic mode \eqref{eq:wigner_q} for $\delta \alpha = -0.2$ and different harmonic cutoff, (a) $N = 5$, (b) $N = 11$, (c) $N = 21$ and (d) $N = 101$.}
      \label{fig:wignerq}
\end{figure}

The Wigner function of the coherent state superposition of the $q$-th harmonic is shown in Fig. \ref{fig:wignerq}, and shows the opposite behavior as the coherent state superposition of the fundamental mode (compare Fig. \ref{fig:wigner_decoherence}). While the non-classical signatures of the Wigner function of the fundamental mode gets protected for increasing harmonic cutoff, the coherent state superposition of the harmonic mode simply show a Gaussian shape for large $N$ corresponding to a classical Wigner function. In contrast, for smaller harmonic cutoff the Wigner function of the coherent state superposition of the $q$-th harmonic reveal its non-classical characteristics by virtue of the negative values.

\section{\label{sec:entanglement_expressions}Reduced density matrices and linear entropy}

Here we provide the exact reduced density matrices and the corresponding linear entropy measures used in the main text for the entanglement measure. 

For the single color high harmonic generation procedure, we consider the state in Eq.(2) of the main manuscript.
The reduced state of the fundamental mode after tracing over the harmonic degrees of freedom $\rho_{q=1} = \Tr_{q>1} (\dyad{\psi})$ reads 
\begin{align}
\label{eq:densitymatrix}
\rho_{1} = & \mathcal{N}^2 \left[ \dyad{\alpha + \delta \alpha} + e^{- \abs{\delta \alpha}^2}  e^{- \Omega} \dyad{\alpha}  - e^{- \frac{1}{2} \abs{\delta \alpha}^2}  e^{- \Omega} \Bigl( e^{-i \varphi}  \dyad{\alpha}{\alpha + \delta \alpha} + e^{i \varphi}  \dyad{\alpha + \delta \alpha}{\alpha} \Bigr) \right],
\end{align}
with normalization $\mathcal{N} = \left[1 - \exp(- \sum_{q = 1} \chi_q )\right]^{-1/2}$. 
And hence the linear entropy $S_{lin} = 1- \Tr (\rho^2)$ is given by 
\begin{align}
\label{eq:lin_entropy}
S_{lin}^{1} =  1 - \mathcal{N}^4 \left[ 1 - \left( e^{- 2\abs{\delta \alpha}^2} - 2 e^{-\abs{\delta \alpha}^2 } \right) \Bigl( e^{- 2 \Omega} - 2 e^{- \Omega} \Bigr) \right].
\end{align}

Furthermore, we compute the entanglement witness between $n$ harmonic modes $q \in \{ 2, \, ... , \, n+1 \}$ and the residual field modes $q^\prime \in \{ 1, \, n+2, \, ... , \, N \}$  (including the fundamental mode $q=1$). The reduced density matrix of the $n$ harmonic modes is given by $\rho_{nq } = \Tr_{q^\prime \neq q} (\dyad{\psi})$ and reads
\begin{align}
\rho_{nq} = & \mathcal{N}^2 \left[ \bigotimes_{q=2}^{n+1} \dyad{\chi_q} + e^{- \abs{\delta \alpha}^2} e^{- \Omega^\prime_{N-n}} \bigotimes_{q=2}^{n+1} \dyad{0_q}  - e^{- \abs{\delta \alpha}^2} e^{- \Omega^\prime_{N-n}} \Bigl( \bigotimes_{q=2}^{n+1} e^{- \frac{1}{2} \abs{\chi_q}^2} \dyad{\chi_q}{0_q} + \bigotimes_{q=2}^{n+1} e^{- \frac{1}{2} \abs{\chi_q}^2} \dyad{0_q}{\chi_q} \Bigr) \right],
\end{align}
where $\Omega^\prime_{N-n} = \sum_{n+2}^N \abs{\chi_{q^\prime}}^2 = \abs{\chi}^2 \frac{N-n}{2}$, and the corresponding linear entropy is given by 
\begin{align}
\label{eq:lin_entropy_q}
S_{lin}^{nq} =& 1 - \mathcal{N}^4 \left[1- e^{- \abs{\delta \alpha}^2} e^{- \Omega}  \left( 2 - e^{- \abs{\delta \alpha}^2} e^{- \Omega^\prime_{N-n}} \right)  \right.  \left. \left( 2- e^{- \Omega^\prime_n } \right) \right], 
\end{align}
where we have the relation $\Omega^\prime_{N-n} = \Omega (N-n) / (N-1)$ and $\Omega^\prime_n = \abs{\chi}^2 \frac{n-1}{2} = \Omega (n-1)/(N-1)$ for equal and only odd high harmonic amplitudes. 

For the two-color high harmonic generation procedure, which leads to the generation of large amplitude entangled coherent states of the two driving fields, we are interested in the associated entanglement. 
To quantify the entanglement between the two driving field modes we proceed by tracing over the second $2 \omega$ driving field and compute the linear entropy of the remaining state from Eq.(11) of the main manuscript. The reduced density matrix of the first driving mode reads 
\begin{align}
\rho_\omega = & \mathcal{N}_{2 \omega}^2 \left[ \dyad{\alpha_1 + \delta \alpha_1} + e^{- \Delta} e^{- 2 \bar\Omega} \dyad{\alpha_1} -  e^{- \frac{1}{2} ( \Delta + \abs{\delta \alpha_2}^2)} e^{- \bar \Omega} \left[ e^{- i \varphi_1} \dyad{\alpha_1}{\alpha_1 + \delta \alpha_1}   + e^{i \varphi_1} \dyad{\alpha_1 + \delta \alpha_1}{\alpha_1} \right]\right], 
\end{align}
where $\Delta = \abs{\delta \alpha_1}^2 + \abs{\delta \alpha_2}^2$, $\varphi_i = \operatorname{Im}(\alpha_i \delta \alpha_i)$ and $\bar \Omega$ is given by \eqref{eq:decoherence_2color}. 
And accordingly the linear entropy is given by 
\begin{align}
\label{eq:two_color_entropy}
S_{lin}^\prime = & 1 - \mathcal{N}_{2\omega}^4 \left[ 1 - 2 e^{-\Delta} e^{- \bar \Omega} \left[ 2 - e^{- \bar \Omega} \left( e^{- \abs{\delta \alpha_1}^2}  +  e^{- \abs{\delta \alpha_2}^2} \right) \right]  + e^{- 2 \Delta} e^{- 2 \bar \Omega} \left(2 - 4 e^{- \bar \Omega} + e^{- 2 \bar \Omega}  \right)  \right],
\end{align}
with normalization $\mathcal{N}_{2\omega} =[1+ e^{- \Delta} (e^{- 2 \bar \Omega} - 2 e^{- \bar \Omega})]^{-1/2}$.

\bibliographystyle{plain}
\bibliography{references}